\documentclass[final,5p,twocolumn]{elsarticle}

\usepackage{amssymb}
\usepackage{adjustbox}
\usepackage{caption}
\usepackage{subcaption}
\usepackage{algorithm}
\usepackage{algorithmic}
 \usepackage{amsmath}

\usepackage{multirow}

\journal{Elsevier}

\begin{document}

\begin{frontmatter}

\title{Battery and Hydrogen Energy Storage Control in a Smart Energy
Network with Flexible Energy Demand using Deep
Reinforcement Learning}


 %\title{Title\tnoteref{label1}}
%%\tnotetext[inst1]{}
%\author{Name\corref{cor1}\fnref{label2}}
\author{Cephas Samende\corref{cor1}\fnref{inst1}}
 \ead{c.samende@keele.ac.uk}
 %\author[inst1]{}
\cortext[cor1]{Corresponding author:}
\affiliation[inst1]{organization={School of Computer Science and Mathematics},%Department and Organization
            addressline={Keele University}, 
            city={Newcastle},
            %postcode={ST5 5BG}, 
            %state={Staffordshire},
            country={United Kingdom}}


\author[inst1]{Zhong Fan}

 \author[inst2]{Jun Cao}
 \affiliation[inst2]{organization={Environmental Research and Innovation Department, Sustainable Energy Systems Group},
             addressline={Luxembourg Institute of Science and Technology},
            city={Esch-sur-Alzette},
            country={Luxembourg}}

\begin{abstract}
Smart energy networks provide for an effective means to accommodate high penetrations of variable renewable energy sources like solar and wind, which are key for deep decarbonisation of energy production. However, given the variability of the renewables as well as the energy demand, it is imperative to develop effective control and energy storage schemes to manage the variable energy generation and achieve desired system economics and environmental goals. In this paper, we introduce a hybrid energy storage system composed of battery and hydrogen energy storage to handle the uncertainties related to  electricity prices, renewable energy production and consumption. We aim to improve renewable energy utilisation and minimise energy costs and carbon emissions while ensuring energy reliability and stability within the network. To achieve this, we propose a multi-agent deep deterministic policy gradient approach, which is a deep reinforcement learning-based control strategy to optimise the scheduling of the hybrid energy storage system and energy demand in real-time. The proposed approach is model-free and does not require explicit knowledge and rigorous mathematical models of the smart energy network environment. Simulation results based on real-world data show that: (i) integration and optimised operation of the hybrid energy storage system and energy demand reduces carbon emissions by 78.69\%, improves cost savings by 23.5\% and renewable energy utilisation by over 13.2\% compared to other baseline models and (ii) the proposed algorithm outperforms the state-of-the-art self-learning algorithms like deep-Q network.
\end{abstract}




\end{frontmatter}



\section{Introduction}
Globally, the energy system is responsible for about 73.2\% of greenhouse gas emissions \cite{ritchie2020}. Deep reductions of greenhouse gas emissions in the energy system  are key for achieving a net-zero greenhouse gas future to limit the rise in global temperatures to 1.5°C and to prevent the daunting effects of climate change \cite{allen2018summary}.  In response, the global energy system is undergoing an energy transition from the traditional high-carbon to a low or zero carbon energy system, mainly driven by enabling technologies like internet of things \cite{fuller2020digital} and high penetration of variable renewable energy sources (RES) like solar and wind \cite{bouckaert2021net}. Although RES are key for delivering a decarbonised energy system  which is reliable, affordable and fair for all, the uncertainties related to their energy generation as well as energy consumption remains a significant barrier, which are unlike the traditional high-carbon system with dispatchable sources \cite{paul2010role}.

Smart energy networks (SEN) (also known as micro-grids), which are autonomous local energy systems equipped with RES, energy storage system (ESS) as well as various types of loads are an effective means of integrating and managing high penetrations of variable RES in the energy system \cite{harrold2022renewable}.  Given the uncertainties with RES energy generation as well as the energy demand, ESSs such as a battery energy storage system (BESS) have proved to play a crucial role in managing the uncertainties while providing reliable energy services to the network \cite{arbabzadeh2019role}. However, due to low capacity density, BESS cannot be used to manage at-scale penetration of variable RES \cite{desportes2021deep}. 

Hydrogen energy storage systems (HESS) are emerging as a promising high capacity density energy storage carriers to support high penetrations of RES. This is mainly due to falling costs for electricity from RES and improved electrolyzer technologies whose costs have fallen by more than  60\% since 2010 \cite{qazi2022future}. During periods of over generation from the RESs, HESSs convert the excess power into hydrogen gas, which can be stored in a tank. The stored hydrogen can be sold externally as fuel such as for use in fuel-cell hybrid electric vehicles \cite{correa2017performance} or converted into power during periods of minimum generation from the RES to complement other ESSs such as the BESS. 
%\cite{9735383, 8858041, qazi2022future}.
%https://www.irena.org/-/media/Files/IRENA/Agency/Publication/2020/Nov/IRENA_Green_Hydrogen_breakthrough_2021.pdf?la=en&hash=40FA5B8AD7AB1666EECBDE30EF458C45EE5A0AA6

The SEN combines power engineering with information technology to manage the generation, storage and consumption to  provide a number of technical and economic benefits such as increased utilization of RES in the network, reduced energy losses and costs, increased power quality, and enhanced system stability \cite{harrold2020battery}. However, this requires an effective smart control strategy to optimise the operation of the ESSs and energy demand to achieve the desired system economics and environmental outcomes.

Many studies have proposed control strategies that optimise the operation of ESSs to minimise the utilization costs \cite{vivas2020suitable, cau2014energy, enayati2022optimal, hassanzadehfard2020design}. Others have  proposed control models for optimal sizing and planning of the micro-grid \cite{castaneda2013sizing,liu2021optimal, pan2020optimal}. Other studies have modelled the optimal energy sharing  in the micro-grid \cite{tao2020integrated}.   Despite a rich history, the proposed control approaches are model-based, in which they require the explicit knowledge and rigorous mathematical models of the micro-grid to capture complex real-world dynamics. Model errors and model complexity  makes them difficult to apply and to optimise the ESSs  in real-time. Moreover, even if an accurate and efficient model without errors exists, it is often a cumbersome and fallible process to develop and maintain the control approaches in situations where uncertainties of the micro-grid are dynamic in nature \cite{nakabi2021deep}. 

In this paper, we propose a model-free control strategy based on reinforcement learning (RL), a machine learning paradigm, in which an agent learns the optimal control policy by interacting with the SEN environment \cite{sutton2018reinforcement}. Through trial and error, the agent selects control actions that maximise a cumulative reward (e.g. revenue) based on its observation of the environment. Unlike the model-based optimisation approaches, model-free-based algorithms do not require explicit knowledge and rigorous mathematical models
of the environment, making them capable of determining optimal control actions in real-time even for complex control problems like peer-to-peer energy trading \cite{samende2022multi}. Further, artificial neural networks can be combined with RL to form deep reinforcement learning (DRL), making model-free approaches capable of handling even more complex control problems \cite{mnih2015human}. Examples of commonly used DRL-based algorithms are value-based algorithms such
as Deep Q-networks (DQN) \cite{mnih2015human} and policy-based algorithms such as deep
deterministic policy gradient (DDPG) \cite{lillicrap2015continuous}.

\subsection{Related Works}
Application of DRL approaches for managing SENs has increased in the past decade. However, much progress has been for SENs having a single ESS (e.g. BESS) \cite{harrold2020battery, nakabi2021deep, wan2021data,8742669, sang2022deep, 9585298, mbuwir2020reinforcement}. With declining costs of RES, additional ESSs like HESS are expected in SENs to provide  additional
system flexibility and storage to support
further deployment of RES. In this case, control approaches that can effectively schedule the hybrid operation of BESS and HESS become imperative.

Recent studies on optimised control of SENs having multiple ESSs like a hybrid of BESS and HESS are proposed in \cite{desportes2021deep, chen2022optimal, zhu2022optimal, tomin2019deep, yu2021optimal}. In \cite{desportes2021deep, chen2022optimal}, a DDPG-based algorithm is proposed to minimise building carbon emissions in a SEN which includes BESS, HESS and constant building loads. Similarly, operating costs are minimised in \cite{zhu2022optimal} using DDPG and in \cite{tomin2019deep} using DQN. However, these studies use a single control agent to manage the multiple ESSs. Energy management of a SEN is usually a multi-agent problem where an action of one agent affects the actions of others, making the SEN environment to be non-stationary from an agent’s perspective \cite{samende2022multi}. Single agents have been found to perform poorly in non-stationary environments \cite{lowe2017multi}.   

A multi-agent based control approach for optimal operation of a hydrogen based multi-energy systems is proposed in \cite{yu2021optimal}. Despite the approach addressing the drawbacks of the single agent, flexibility of the electrical load is not investigated.  With the introduction of flexible loads like heat pumps which run on electricity in SENs \cite{heat2019delivering}, the dynamics of the electrical load is expected to change the technical-economics and the environmental impacts of the SEN.

Compared with the existing works, we investigate a SEN that has a BESS, HESS and a schedulable energy demand. We explore the energy cost and carbon emission minimisation problem of a such a SEN while capturing the time-coupled storage dynamics of the BESS and the HESS, as well as the uncertainties related to RES, varying energy prices and the flexible demand. A multi-agent deep deterministic policy gradient (MADDPG) algorithm is  developed to reduce the system cost and carbon emissions, and to improve the utilisation of RES while addressing the drawbacks of a single agent in a non-stationary environment. To the authors’ knowledge, this study is the first to comprehensively apply the MADDPG algorithm to optimally schedule operation of the hybrid BESS and HESS as well as the energy demand in a SEN.
\subsection{Contributions}
The main contributions of this paper are on the following aspects:
\begin{itemize}
    \item  We formulate the system cost minimisation problem of the SEN, complete with BESS, HESS, flexible demand, solar and wind generation as well as dynamic energy pricing as a function of energy costs and carbon emissions cost.  The system cost minimisation problem is then reformulated as a continuous action based Markov game with unknown probability to adequately obtain the optimal energy control policies without explicitly estimating the underlying model of the SEN and relying on future information.
   \item A data-driven self-learning based MADDPG algorithm that outperforms a model-based solution and other DRL-based algorithms used as a benchmark is proposed to solve the Markov game in real-time. This also includes the use of a novel real-world generation and consumption data set collected from the Smart Energy Network Demonstrator (SEND) project at Keele University\footnote{https://www.keele.ac.uk/business/businesssupport/smartenergy/}.
   \item We carry out a simulation analysis of a SEN model for five different scenarios to demonstrate the benefits of integrating a hybrid of BESS and HESS  as well as  scheduling  the energy demand in  the network.
    \item Simulation results based on SEND data show that the proposed algorithm can increase cost savings and reduce carbon emission by 41.33\% and 56.3\% respectively compared with other bench-marking algorithms and baseline models.

\end{itemize}

The rest of the paper is organized as follows. Description of the SEN environment is presented in Section \ref{sec__1}. Formulation of the optimisation problem is given in Section \ref{problem_formulation}. A brief background to RL and the description of the proposed self-learning algorithm is presented in Section \ref{sec:MADRL}. Simulation results are provided in Section \ref{case study}, with conclusions presented in Section \ref{conclusion}.

\section{Smart Energy Network}\label{sec__1}

\begin{figure}[t!]
    \centering
    \includegraphics[width=0.48\textwidth]{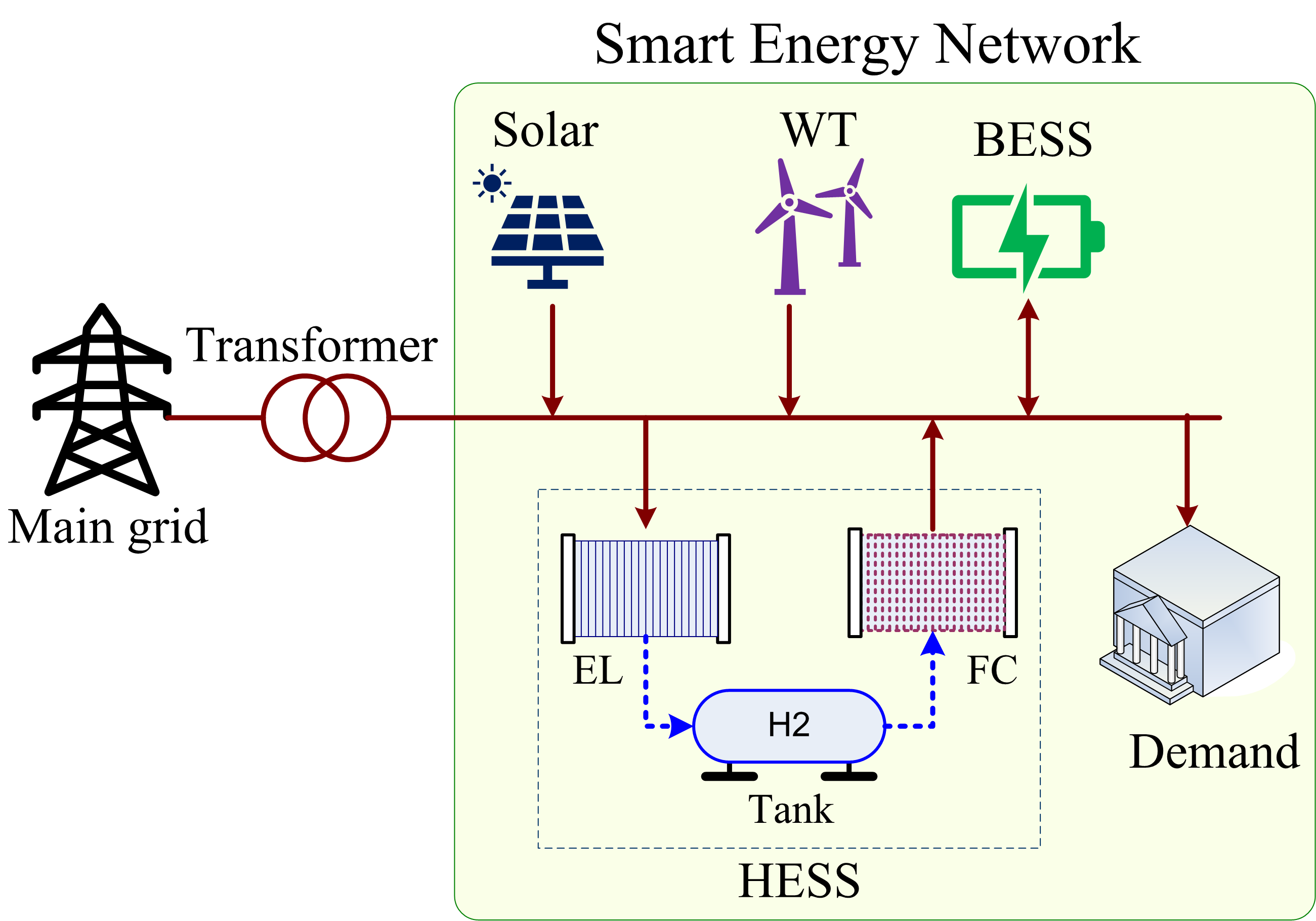}
    \caption{Basic structure of the grid-connected smart energy network, which consists of solar, wind turbines (WT), flexible energy demand, battery energy storage system (BESS), and hydrogen energy storage system (HESS). The HESS consists of three main components, namely electrolyzer (EL), storage tank and fuel-cell (FC). Solid lines represent electricity flow. Dotted lines represent flow of hydrogen gas.}
    \label{fig:SEN}
    \vspace{-\baselineskip}
\end{figure}

The SEN considered in this paper is a grid-connected micro-grid with RES (solar and wind turbines), hybrid energy storage system (BESS and HESS) and the electrical energy demand as shown in Fig. \ref{fig:SEN}. The aggregated electrical demand from the building(s) is considered to be a price responsive demand i.e., the demand can be reduced based on electricity price variations or shifted from the expensive price time slots to the cheap price time slots. At every time slot \(t\), solar and wind turbines provide energy to meet the energy demand. Any excess generation is either used to charge the BESS and/or converted into hydrogen by the electrolyzer or exported to the main grid at a feed-in tariff \(\pi_t\). In the events that energy generated from solar and wind turbines is insufficient to meet the energy demand, the deficit energy is either supplied by the BESS and/or fuel-cell or imported from the main grid at a time-of-use (ToU) tariff \(\lambda_t\).

 In the following sub-sections, we present  models of solar, wind, BESS, HESS (i.e., electrolyzer, tank and fuel cell) and flexible demand adopted in this paper.
\subsection{PV and Wind Turbine Model}
 Instead of using mathematical equations to model the solar and wind turbine, we use real energy production data from the solar and the wind turbine as these are un-dispatchable under normal SEN operating conditions. Thus, at every time step \(t\), power generated from solar and wind turbine is modelled as \(P_{pv,t}\) and  \(P_{w,t}\) respectively.  

\subsection{BESS Model}
The key property of a BESS is the amount of energy it can store at time \(t\).
Let \(P_{c,t}\) and \(P_{d,t}\) be the charging and discharging power of the BESS respectively. The BESS energy dynamics during charging and discharging operation can be modelled as follows \cite{9454450}
\begin{equation}
    E_{t+1}^b = E_t^b + \Big(\eta_{c,t}P_{c,t} - \frac{P_{d,t}}{\eta_{d,t}}\Big)\Delta t,\;\;\;\;\forall t    \label{eq:batE}
\end{equation}
where \(\eta_{c,t} \in (0,1]\) and \(\eta_{d,t} \in (0,1]\) are dynamic BESS charge and discharge efficiency as calculated in \cite{9534877} respectively, \(E_t^b\) is the BESS energy (kWh) and \(\Delta t\) is the duration of BESS charge or discharge.

The BESS charge level is limited by the storage capacity of the
BESS as
\begin{equation}
    E_{min}\le E_t^b \le E_{max} \label{ineq_B}
\end{equation}
where \(E_{min}\) and \(E_{max}\) are lower and upper boundaries of the BESS charge level.

To avoid charging and discharging the BESS at the same time, we have
\begin{equation}
    P_{c,t}\cdot P_{d,t} = 0,\;\;\;\;\forall t \label{dot_b}
\end{equation}
That is, at any particular time \(t\), either \(P_{c,t}\) or \(P_{d,t}\) is zero.

Further, the charging and discharging power is limited by maximum battery terminal power \(P_{max}\) as specified by manufacturers  as
\begin{equation}
  0\leq P_{c,t}, P_{d,t}\leq P_{max},\;\;\;\;\forall t\label{ineq_pc}
\end{equation}

During operation, the BESS wear cannot be avoided due to repeated BESS charge and discharge processes. The wear cost can have a great impact on the economics of the SEN.  The empirical wear cost of the BESS can be expressed as \cite{han2014practical}
\begin{equation}
    C_{BESS}^t = \frac{C_b^{ca}|E_t^b|}{L_c\times2\times\text{DoD}\times E_{nom}\times (\eta_{c,t} \times \eta_{d,t})^2}
\end{equation}
where \(E_{nom}\) is the BESS nominal capacity, \(C_b^{ca}\) is the BESS capital cost, DoD is the depth of discharge at which the BESS is cycled, and  \(L_c\) is the BESS life cycle.

\subsection{HESS Model}
In addition to the BESS, a HESS is considered in this study as a long-term energy storage unit. The HESS mainly consists of an electrolyzer (EL), hydrogen storage tank (HT) and fuel cell (FC) as shown in Fig. \ref{fig:SEN}. The electrolyzer uses the excess electrical energy from the RESs to produce hydrogen. The produced hydrogen gas is stored in the hydrogen storage tank and later used by the fuel cell to produce electricity whenever there is a deficit energy generation in the SEN. 

Dynamics of hydrogen in the tank associated with the generation and consumption of hydrogen by the electrolyzer and fuel cell respectively is modelled  as follows \cite{vivas2020suitable}
\begin{equation}
  H_{t+1} = H_t + \Big(r_{el,t}P_{el,t} - \frac{P_{fc,t}}{r_{fc,t}}\Big) \Delta t ,\;\;\;\;\forall t \label{hess_eq}
\end{equation}
where \(P_{el,t}\) and \(P_{fc,t}\)  are the electrolyzer power input and fuel cell output power respectively, \(H_t\) (in Nm\(^3\)) is hydrogen gas level in the tank, \(r_{el,t}\) (in Nm\(^3\)/kWh) and \(r_{fc,t}\) (in kWh/Nm\(^3\)) are the hydrogen generation and consumption ratios associated with the electrolyzer  and fuel cell respectively.

The hydrogen level is limited by the storage capacity of the tank as
\begin{equation}
  H_{min}\leq  H_t\leq H_{max},\;\;\;\;\forall t \label{ineq_H}
\end{equation}
where \(H_{min}\) and \(H_{max}\) are the lower and upper boundaries imposed on the hydrogen level in the tank.

As the electrolyzer and the fuel cell cannot operate at the same time, we have
\begin{equation}
  P_{el,t}\cdot P_{fc,t} = 0,\;\;\;\;\forall t\label{dot_h}
\end{equation}
Furthermore, power consumption and power generation respectively associated with the electrolyzer and fuel cell is restricted to their rated values as
\begin{align}
  0\leq P_{el,t}&\leq P_{max}^{el},\;\;\;\;\forall t \label{ineq_pe}\\
  0\leq P_{fc,t}&\leq P_{max}^{fc},\;\;\;\;\forall t\label{ineq_pf}
\end{align}
where \(P_{max}^{el}\) and \(P_{max}^{fc}\) are the rated power values of the electrolyzer and fuel cell respectively.

If the HESS is selected to store the excess energy, the cost of producing hydrogen through the electrolyzer and later becoming fuel cell energy  is given as \cite{dufo2007optimization}
\begin{equation}
    C_t^{el-fc} = \frac{(C_{el}^{ca}/L_{el}+C_{el}^{om})+(C_{fc}^{ca}/L_{fc}+C_{fc}^{om})}{\eta_{fc,t}\eta_{el,t}}
\end{equation}
where \(C_{el}^{ca}\) and \(C_{fc}^{ca}\) are electrolyzer and fuel cell capital costs, \(C_{el}^{om}\) and \(C_{fc}^{om}\) are the operation and maintenance  costs of the electrolyzer and the fuel cell, \(\eta_{el,t}\) and \(\eta_{fc,t}\) are the electrolyzer and fuel cell efficiencies, \(L_{el}\) and \(L_{fc}\) are the electrolyzer and the fuel cell lifetimes respectively.

The cost of meeting the deficit energy using the fuel cell with the  hydrogen stored in the tank as fuel is given as \cite{cau2014energy}
\begin{align}
     C_t^{fc} = &\frac{C_{fc}^{ca}}{L_{fc}} + C_{fc}^{om}
\end{align}
The total cost of operating the HESS at time \(t\) can be expressed as follows
 \begin{align}
 C_{HESS}^t = 
 \begin{cases}
     C_t^{el-fc},\;\;\;\text{if}\;\;P_{el,t} > 0 \\
     C_t^{fc},\;\;\;\text{if} \;\;P_{fc,t} > 0 \\
     0,\;\;\;\text{otherwise}
\end{cases}
\end{align}

\subsection{Load Model}
We assume that the total energy demand of the SEN has a certain proportion of flexible energy demand that can be reduced or shifted in time due to the energy price. Thus, at every time \(t\), the actual demand may deviate from the expected total energy demand. Let the total energy demand before energy reduction be \(D_t\) and the actual energy demand after reduction be \(d_t\). Then the energy reduction \(\Delta d_t\) can be expressed as
\begin{align}
     \Delta d_t = D_t - d_t\;\;\;\;\forall t \label{dem_eq}
\end{align}
As reducing the energy demand inconveniences the energy users, the \(\Delta d_t\) can be constrained as follows
\begin{align}
    0\le  \Delta d_t \le \zeta D_t\;\;\;\;\forall t l\label{dem_ch}
\end{align}
where \(\zeta\) (e.g., \(\zeta = 30\%\)) is a constant factor that specifies the maximum percentage of original demand that can be reduced.

\begin{figure}[t!]
    \centering
    \includegraphics[width=0.48\textwidth]{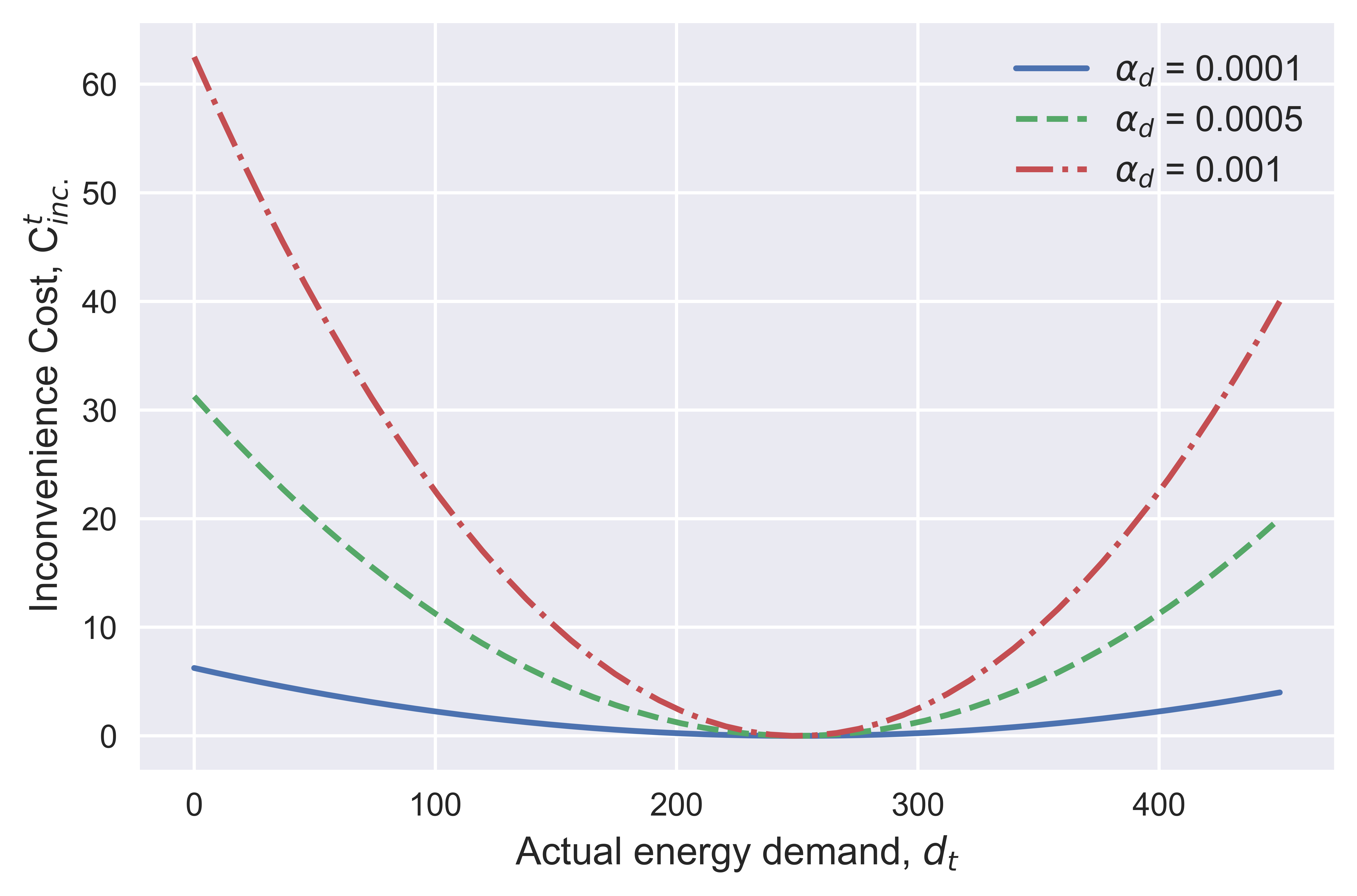}
    \caption{Impact of \(\alpha_d\) parameter on the inconvenience cost of the energy demand, when \(D_t = 250\) kW and when \(d_t\) takes values from 0 to 450 kW.}
    \label{fig_inc}
    \vspace{-\baselineskip}
\end{figure}

The inconvenience cost for reducing the energy demand  can be estimated using a convex function as follows
\begin{align}
    C_{inc.}^t = \alpha_d\Big(d_t - D_t\Big)^2\;\;\;\;\forall t \label{inc_cost}
\end{align}
where \(\alpha_d\) is a small positive number that quantifies  the amount of flexibility  to reduce the energy demand as shown in Fig. \ref{fig_inc}. A lower value of \(\alpha_d\) indicates that less attention is paid to the inconvenience cost and a larger share of the energy demand can be reduced to minimise the energy costs. A higher value of \(\alpha_d\) indicates that high attention is paid to the inconvenience cost and the energy demand can be hardly reduced to minimise the energy costs.

\subsection{SEN Energy Balance Model}
 Local RES generation and demand in the SEN must be matched at all times for stability of the energy system. Any energy deficit and excess  must be imported and exported to the main grid respectively. 
 
 The power import and export at time \(t\) can be expressed as 
  \begin{align}
     P_{g,t} = d_{t} + P_{c,t} + P_{el,t} - P_{pv,t} - P_{w,t} - P_{d,t} - P_{fc,t}  \label{eq_balance}
 \end{align}
 where \(P_{g,t}\) is power import if \(P_{g,t} > 0\) and power export otherwise. We assume that the SEN is well sized and that \(P_{g,t}\) is always within the allowed export and import power limits.
 
 Let \(\pi_t\) and \(\lambda_t\) be the export and import grid prices at time \(t\) respectively.  As grid electricity is  the major source of carbon emissions, the cost of utilising the main grid to meet the supply-demand balance in the SEN is the sum of both the energy cost and the environmental cost due to carbon emissions as follows
 \begin{align}
 C_{grid}^t =\Delta t 
 \begin{cases}
     \lambda_tP_{g,t} + \mu_{c}P_{g,t},\;\;\;\text{if}\;\;P_{g,t}\geq 0 \\
      -\pi_t|P_{g,t}|,\;\;\;\text{otherwise} 
\end{cases}
\end{align}
where \(\mu_{c}\in [0,1]\) is the carbon emission conversion factor of grid electricity.

\section{Problem Formulation}\label{problem_formulation}
The key  challenge in operating the SEN is with how to optimally schedule the operation of the BESS, HESS and the flexible energy demand to minimise energy costs and carbon emissions as well as to increase renewable energy utilisation.  The operating costs associated with PV and wind generation are neglected for being comparatively smaller than those for energy storage units and energy demand \cite{vivas2020suitable}.

\subsection{Problem Formulation}
As the only controllable assets in the SEN considered in this paper are the BESS, HESS and the flexible energy demand, the control variables can be denoted as a vector \(\mathbf{v}_t = \{P_{c,t}, P_{d,t}, P_{el,t},P_{fc,t}, \Delta d_t\}\). The \(P_{g,t}\) can be obtained according to (\ref{eq_balance}). We formulate a system overall cost-minimizing problem as a function of the energy costs and the environmental cost as follows
\begin{equation*}
\mathbf{P1:}\quad
\begin{aligned}
\min_{\mathbf{v}_t}:  &\sum_{t=1}^{T}\Big( C_{BESS}^t +  C_{HESS}^t +  C_{inc.}^t +  C_{grid}^t\Big)\\
\textrm{s.t.:}  & (\ref{eq:batE})-(\ref{ineq_pc})\;\&\;(\ref{hess_eq})-(\ref{ineq_pf})\;\&\: (\ref{dem_eq}),(\ref{dem_ch}),(\ref{eq_balance}) 
\end{aligned}\label{opt}
\end{equation*}
Solving this optimisation problem using model-based optimisation approaches suffers from three main challenges, namely uncertainties of parameters, information and dimension  challenges. The uncertainties are related to RES, energy price as well as energy demand, which makes it  difficult to directly solve the optimisation problem without statistical information of the system. As expressed in (\ref{eq:batE}) and (\ref{hess_eq}), control of the BESS and HESS is time-coupled and actions taken at time \(t\)  have an effect on future actions to be taken at time \(t+1\). Thus, for optimal scheduling, the control policies should also consider the future `unknown' information of the BESS and the HESS. Moreover, the control actions  of the BESS and the HESS are continuous in nature and bounded as given in (\ref{ineq_pc}), (\ref{ineq_pe}), (\ref{ineq_pf}), which increases the dimension of the control problem.  

In the following sub-sections, we overcome these challenges by first, re-formulating the optimisation problem as a continuous action Markov-game and later solving it using a self-learning  algorithm.

\subsection{Markov-Game Formulation}
 We reformulate \textbf{P1} as a Markov decision process (MDP) which consists of a state space \(\mathcal{S}\), an action space \(\mathcal{A}\), a reward function \(\mathcal{R}\), a discount factor \(\gamma\) and a transition probability function \(\mathcal{P}\) as follows:

\subsubsection{\textbf{State Space}} The state space \(\mathcal{S}\) represents the collection  of all  the state variables of the SEN at every time slot \(t\) including RES variables (\(P_{pv,t}\;\&\;P_{w,t}\)), energy prices (\(\pi_t\;\&\;\lambda_t\)), energy demand \(D_t\) and state of the ESSs (\(E_t^b\;\&\:H_t\)). Thus, at time slot \(t\), the state of the system is given as
\begin{equation}
    s_t = \Big(P_{pv,t}, P_{w,t}, E_{b,t}, H_{t}, D_{n,t},\pi_t,\lambda_t\Big),\;\;s_t  \in \mathcal{S} 
\end{equation}

\subsubsection{\textbf{Action Space}}  The action space denotes the collection of all actions \(\{P_{c,t}, P_{d,t}, P_{el,t}, P_{fc,t}, \Delta d_t\}\), which are the decision values of \textbf{P1} taken by the agents to produce the next state \(s_{t+1}\) according to the state transition function \(\mathcal{P}\). To reduce the size of the action space, action variables for each storage system can be combined into one action. With reference to (\ref{dot_b}), the BESS action variables \(\{P_{c,t},  P_{d,t}\}\) can be combined into one action \(P_{b,t}\) so that during charging (i.e. \(P_{b,t} < 0\)), \(P_{c,t} = |P_{b,t}| \;\&\; P_{d,t} = 0\). Otherwise, \(P_{d,t} = P_{b,t}  \;\&\; P_{c,t} = 0\). Similarly, the HESS action variables \(\{P_{el,t}, P_{fc,t}\}\) can be combined into one action \(P_{h,t}\). During electrolysis, (i.e. \(P_{h,t} < 0\)) , \(P_{el,t} = |P_{h,t}| \;\&\; P_{fc,t} = 0\). Otherwise, \(P_{fc,t} = P_{h,t} \;\&\; P_{el,t} = 0\). Thus, at time \(t\), the control actions of the SEN reduces to
\begin{equation}
   a_t  = \Big(P_{b,t}, P_{h,t}, \Delta d_t\Big),\;\;a_t \in \mathcal{A}
\end{equation}
The action values are bounded according to their respectively boundaries given by (\ref{ineq_pc}),  (\ref{ineq_pe}), (\ref{ineq_pf}) and (\ref{dem_ch}).

\subsubsection{\textbf{Reward Space}} The collection of all the rewards received by the agents after interacting with the environment forms the reward space \(\mathcal{R}\). The reward is used to evaluate the performance of the agent based on the actions taken and the state of the SEN observed by the agents at that particular time.  The first part of the reward is the total energy cost and environmental cost of the SEN
\begin{equation}
    r_t^{(1)} = -\Big( C_{BESS}^t +  C_{HESS}^t +  C_{inc.}^t +  C_{grid}^t\Big)
\end{equation}
As constraints given in (\ref{ineq_B}) and (\ref{ineq_H}) should always be satisfied, the second part of the reward is a penalty for violating the constraints as follows
\begin{align}
r_t^{(2)} =- 
\begin{cases}
\mathcal{K},\;\;\;\text{if}\;(\ref{ineq_B})\;\text{or}\; (\ref{ineq_H})\;\text{is violated}\\
 0,\;\;\;\;\text{otherwise}
\end{cases}
\end{align}
where \(\mathcal{K}\) is a predetermined large number, e.g. \(K = 20\).

The total reward received by the agent after interacting with the environment is therefore expressed as
\begin{equation}
    r_t = r_t^{(1)} + r_t^{(2)},\;\;r_t \in \mathcal{R}
\end{equation}
	
The goal of the agent is to maximize its own expected reward \(R\)
\begin{equation}
    R = \sum_{t=0}^{T}\gamma^tr_t\label{cumm_reward___}
\end{equation}
where \(T\) is the time horizon and  \(\gamma\) is a discount factor, which helps the agent to focus the policy by caring more about obtaining the rewards quickly.

As electricity prices, RES energy generation and demand are volatile in nature, it is generally impossible to obtain with certainty the state transition probability function \(\mathcal{P}\)  required to derive an optimal policy \(\pi (s_t|a_t)\) needed to maximize \(R\). To circumvent this difficulty, we propose the use of RL as discussed in Section \ref{sec:MADRL}.

\section{Reinforcement Learning}\label{sec:MADRL}

\subsection{Background}
A RL framework is made up of two main components, namely the environment and agent. The environment denotes the problem to be solved. The agent denotes the learning algorithm. The agent and environment continuously interact with each other \cite{sutton2018reinforcement}.

At every time \(t\), the agent learns for itself  the optimal control policy  \(\pi (s_t|a_t)\) through trial and error by selecting control actions \(a_t\) based on its perceived state \(s_t\) of the environment. In return, the agent receives a reward \(r_t\) and the next state \(s_{t+1}\) from the environment without explicitly having knowledge of the transition probability function \(\mathcal{P}\). The goal of the agent is to improve the policy so as to maximise  the cumulative reward \(R\).
The environment has been described in Section \ref{problem_formulation}. Next, we describe the learning algorithms.
\subsection{Learning Algorithms}
In this section we present three main learning algorithms considered in this paper, namely DQN (a single agent and value-based algorithm), DPPG (a single agent and policy-based algorithm), and the proposed multi-agent DDPG (a multi-agent and policy-based algorithm).
\subsubsection{DQN} The DQN algorithm was developed by Google DeepMind in 2015 \cite{mnih2015human}. It was developed to enhance a classic RL algorithm called Q-Learning \cite{sutton2018reinforcement} through the addition of deep neural networks and a novel technique called experience replay. In Q-learning, the agent learns the best policy  \(\pi (s_t|a_t)\) based on the  notion of an action-value Q-function as \(    Q_{\pi}(s,a) = \mathbb{E}_{\pi}\left[R|s_t = s, a_t = a\right]\).
By exploring the environment, the agent updates the \(Q_{\pi}(s,a)\) estimates using the Bellman Equation as an iterative update as follows:
\begin{equation}
    Q_{i+1}(s_t,a_t) \leftarrow Q_{i}(s_t,a_t) + \alpha h
    \label{eq:bell}
\end{equation}
where \(\alpha \in (0,1]\) is the learning rate and \(h\) is given by:
\begin{equation}
    h = \left[r_t + \gamma\underset{a}{\text{max}}Q_{\pi}(s_{t+1},a) - Q_{i}(s_t,a_t) \right]
\end{equation}
Optimal Q-function \(Q^{*}\) and policy \(\pi^{*}\) is obtained when \(Q_i(s_t, a_t) \rightarrow Q^{*}(s_t, a_t)\) as \(i\rightarrow \infty\).
As the Q-learning represents the Q-function as a table containing values of all combinations of states and actions, it is impractical for most problems. The DQN algorithm addresses such by using a deep neural network with parameters \(\theta\) to estimate the optimal Q-values, i.e.  \(Q(s_t, a_t;\theta) \approx Q^{*}(s_t, a_t)\) by  minimizing the following loss function \(L(\theta)\) at each iteration \(i\):
\begin{equation}
    L_i(\theta_i) =\mathbb{E} \left[\Big(y_i - Q(s_t,a_t;\theta_i)\Big)^2\right]
\end{equation}
where \(y_t  = r_t + \gamma\underset{a}{\text{max}}Q(s_{t+1},a_t;\theta_{i-1}) \) is the target for iteration \(i\).

To improve training and for better data efficiency, at each time step \(t\), an experience, \(e_t = \langle s_t,a_t, r_t, s_{t+1}\rangle \) is stored in a replay buffer \(\mathcal{D}\). During training, the loss and its gradient is then computed using a mini-batch of transitions sampled from the replay buffer. However, DQN and Q-learning both suffer from an overestimation problem as they both use the same action-value to select and evaluate the Q-value function, making them impractical for problems with continuous action spaces.

\subsubsection{DDPG} DDPG algorithm is proposed to  \cite{lillicrap2015continuous} to handle control problems with continuous action spaces, which otherwise are impractical to be handled by Q-lerning and DQN. The DDPG  consists of two independent neural networks: an actor network and a critic network. The actor network is used to approximate the policy \(\pi(s_t|a_t)\).  The input to the actor network is the environment state \(s_t\) and the output is the action \(a_t\). The critic network is used to approximate the Q-function \(Q(s_t, a_t)\) and is only used to train the agent and the network is discarded during the deployment of the agent. The input to the critic network is the concatenation  of the state \(s_t\) and the action \(a_t\) from the actor network and the output is the Q-function \(Q(s_t, a_t)\). 

Similar to the DQN, the DDPG stores an experience, \(e_t = \langle s_t,a_t, r_t, s_{t+1}\rangle \) in a replay buffer \(\mathcal{D}\) at each time step \(t\) to improve training and for better data efficiency. To add more stability to the training, two target neural networks, which are identical to the (original) actor network and (original) critic network are also created. Let the network parameters of the original actor network, original critic network, target actor network, and target critic network be denoted as \(\theta^{\mu}\), \(\theta^{Q}\), \(\theta^{\mu^{'}}\), and \(\theta^{Q^{'}}\)  respectively. 
Before training starts,  $\theta^{\mu}$ and $\theta^{Q}$ are randomly initialized and the \(\theta^{\mu^{'}}\), and \(\theta^{Q^{'}}\) are initialized as $\theta^{\mu^{'}} \leftarrow \theta^{\mu}$ and $\theta^{Q^{'}} \leftarrow \theta^{Q}$. 

To train the original actor and critic networks, a min-batch of \(B\) experiences \(\langle s_t^j,a_t^j, r_t^j, s_{t+1}^j\rangle \Big|_{j=1}^B\), is randomly sampled from  \(\mathcal{D}\), where \(j\in B\) is the sample index. The original critic network parameters $\theta^{Q}$ are updated through gradient descent using the mean-square Bellman error function
\begin{equation}
L\left(\theta^{Q}\right) = \frac{1}{B}\sum\limits_{j=1}^B\Big(y_j - Q\left(s_t^j, a_t^j;\theta^{Q}\right)\Big)^2
\label{critic_loss}
\end{equation} 
where $Q\left(s_t^j, a_t^j;\theta^{Q}\right)$ is the predicted output of the original critic network and \(y_j\) is its target value expressed as
\begin{equation}
y_j = r_t^j + \gamma Q^{'}\left(s_{t+1}^j, \mu^{'} (s_{t+1}^j; \theta^{\mu^{'}});\theta^{Q^{'}}\right)
\end{equation}
where \(\mu^{'}(s_{t+1}^j; \theta^{\mu^{'}})\) is the output (action) from the target actor network and  \(Q^{'}\left(s_{t+1}^j, \mu^{'}(s_{t+1}^j; \theta^{\mu^{'}});\theta^{Q^{'}}\right)\) is the output (Q-value) from the target critic network.

At the same time, parameters of the original actor network are updated by maximising the policy objective function \(J(\theta^{\mu})\)
\begin{equation}
	\nabla_{\theta^{\mu}}J(\theta^{\mu}) = \frac{1}{B}
	\sum\limits_{j=1}^B\nabla_{\theta^{\mu}}\mu\left(s;\theta^\mu\right)
	\nabla_{a}Q\left(s,a;\theta^Q\right)
	\label{policy_grad}	
\end{equation}
where \(s=s_t^j\), \(a = \mu(s_t^j;\theta^\mu)\) is the output (action) from the original actor network and \(Q\left(s,a;\theta^Q\right)\) is the output (Q-value) from the original critic network.

After the parameters of the original actor network and original critic network are updated, the parameters of the two target networks are updated through soft update technique as
\begin{equation}
	\begin{cases}
	\theta^{Q^{'}} \gets \tau \theta^{Q} + \left(1 - \tau\right)\theta^{Q^{'}}\\
	\theta^{\mu^{'}} \gets \tau \theta^{\mu} + \left(1 - \tau\right)\theta^{\mu^{'}}	
	\end{cases}
	\label{target_update}
\end{equation}
where \(\tau\) is the learning rate.

To ensure that the agent explores the environment, a random process \cite{uhlenbeck1930theory} is used to generate a noise \(\mathcal{N}_t\), which is added to every action as follows 
\begin{equation}
	a_t = \mu\left(s_t;\theta^\mu)\right) + \mathcal{N}_t
	\label{action}
\end{equation}
However, as discussed in \cite{lowe2017multi}, the DDPG algorithm performs poorly in non-stationary environments.

\subsection{The Proposed MADDPG Algorithm}\label{sec:MADDPG}
\begin{figure}[t!]
    \centering
    \includegraphics[width=0.48\textwidth]{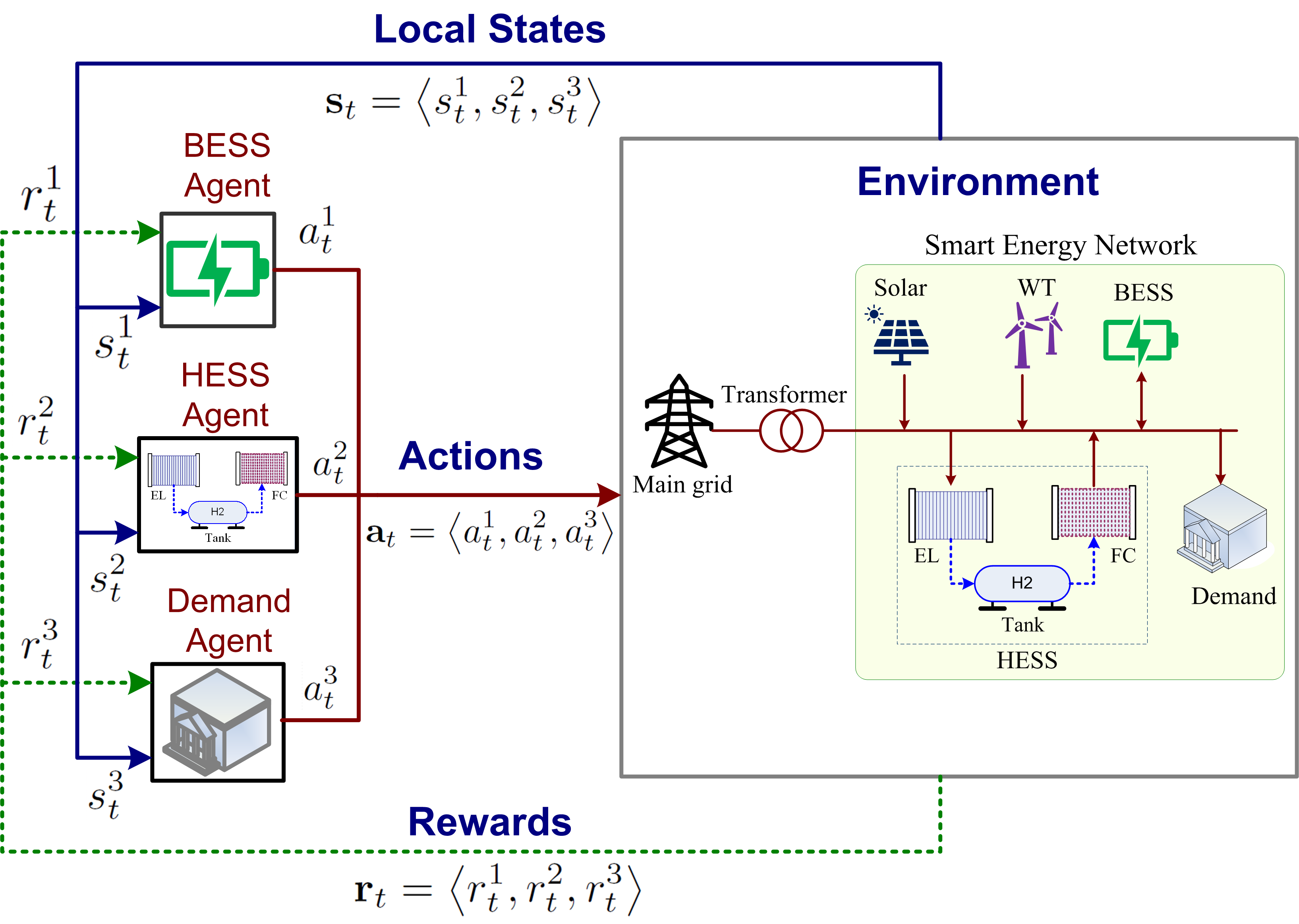}
    \caption{The multi-agent environment structure of the smart energy network.}
    \label{MA_structure}
    \vspace{-\baselineskip}
\end{figure}
Each controllable asset of the SEN (i.e., BESS, HESS and flexible demand) can be considered an agent, making the SEN environment a multi-agent environment as shown in Fig. \ref{MA_structure}. With reference to Section \ref{problem_formulation}, the state and action spaces for each agent can be defined as follows.
 The BESS agent's state and action as \( s_t^1 = (P_{pv,t}, P_{w,t}, E_{b,t}, D_{n,t}, \pi_t,\lambda_t)\) and \(a_t^1  = (P_{b,t})\) respectively. The HESS agent's state and action as \( s_t^2 = (P_{pv,t}, P_{w,t}, D_{n,t}, H_{t}, \pi_t,\lambda_t)\) and \(a_t^2  = ( P_{h,t})\) respectively and the flexible demand agent's state and action as \( s_t^3 = (P_{pv,t}, P_{w,t}, D_{n,t}, \pi_t,\lambda_t)\) and \(a_t^3  = (\Delta d_t)\) respectively. All the agents coordinate to maximise the same cumulative reward function given by (\ref{cumm_reward___}).
\begin{figure*}[t!]
    \centering
    \includegraphics[width=0.8\textwidth]{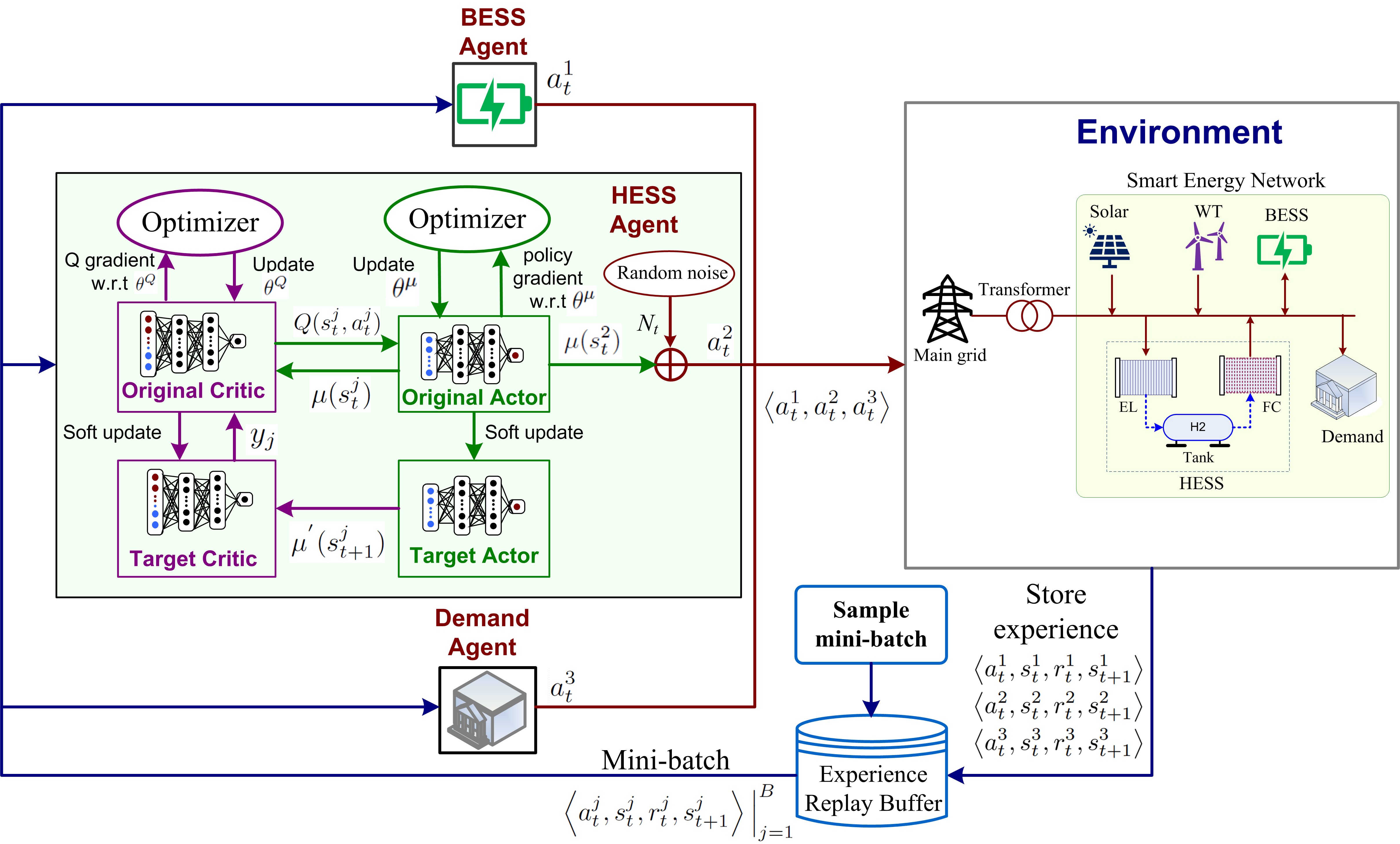}
    \caption{MADDPG structure and training process. The BESS agent and demand agent has the same internal structure as the HESS agent.}
    \label{MA_structure}
    \vspace{-\baselineskip}
\end{figure*}

With the proposed MADDPG algorithm, each agent is modelled as a DDPG agent, where, however, states and actions are shared between the agents during training as shown in Fig. \ref{MA_structure}. During training, the actor network uses only the local state to calculate the  actions while the critic network uses states and actions of all agents in the system in evaluating the local action. As actions of all agents are known by each agent's critic network, the entire environment is stationary during training. During execution, critic networks are removed and only actor networks are used. This means that with MADDPG, training is centralized while execution is decentralized. 

A detailed pseudo-code of the proposed algorithm is given in Algorithm \ref{algorithm}.
\begin{algorithm}
	\caption{MADDPG-based Optimal Control of a SEN}
	\begin{algorithmic}[1]
	    \STATE Initialize shared replay buffer \(\mathcal{D}\) 
		\FOR {each agent \(k = 1,\cdots,3\)}
		\STATE Randomly initialize (original) actor and critic networks with parameters  $\theta^{\mu}$ and $\theta^{Q}$ respectively
		\STATE Initialize (target) actor and critic networks as  $\theta^{\mu^{'}} \leftarrow \theta^{\mu}$ and $\theta^{Q^{'}} \leftarrow \theta^{Q}$ respectively
		\ENDFOR	
		\FOR {each episode \(eps = 1,2,\cdots,M\)}
		\FOR {each agent \(k = 1,\cdots,3\)}
		\STATE Initialize a random process $\mathcal{N}_t$ for exploration
		\STATE Observe initial state \(s_t^k\) from the environment
		\ENDFOR		
		\FOR {each time step $t= 1,2,\cdots,T$}
		\FOR {each agent \(k = 1,\cdots,3\)}
		\STATE Select an action according to (\ref{action})
		\ENDFOR	
		\STATE Execute joint action \(\mathbf{a}_t = \langle a_t^1, a_t^2, a_t^3\rangle\)
		\FOR {each agent \(k = 1,\cdots,3\)}
		\STATE Collect reward \(r_t^k\) and observe state \(s_{t+1}^k\)
		\STATE Store \(\left\langle a_t^k, s_t^k,r_t^k, s_{t+1}^k\right \rangle\) into \(\mathcal{D}\) 
		\STATE Update $s_{t}^k$ $\gets$  $s_{t+1}^k$
		\STATE Randomly sample minibatch of $B$ transitions \(\left\langle a_t^j, s_t^j, r_t^j, s_{t+1}^j \right \rangle\Big|_{j=1}^B\)  from \(\mathcal{D}\) 
		\STATE Update (original) critic network by  (\ref{critic_loss})
		\STATE Update (original) actor network by (\ref{policy_grad})
		\STATE Update target networks by (\ref{target_update})
		\ENDFOR
		\ENDFOR
		\ENDFOR
	\end{algorithmic}
	\label{algorithm}
\end{algorithm}
\section{Simulation Results}\label{case study}

\subsection{Experimental Setup}

In this paper, real-world RES (solar and wind) generation and consumption data which is obtained from the Smart Energy Network Demonstrator (SEND)\footnote{https://www.keele.ac.uk/business/businesssupport/smartenergy/} is used for the simulation studies. We use the UK's time-of-use (ToU) electricity price as grid electricity buying price, which is divided into peak price  \pounds0.234/kWh (4pm-8pm), flat price \pounds0.117/kWh (2pm-4pm \& 8pm-11pm) and the valley price \pounds0.07/kWh (11pm-2pm). The electricity price for selling electricity back to the main grid is a flat price \(\pi_t=\)\pounds0.05/kWh, which is lower than the ToU to avoid any arbitrage behaviour by the BESS and HESS. A carbon emission conversion factor\footnote{https://www.rensmart.com/Calculators/KWH-to-CO2} \(\mu_c = 0.23314\)kgCO\(_2\)/kWh is used to quantify the carbon emissions generated for using electricity from the main grid to meet the energy demand in the SEN. We set the initial BESS state of charge and hydrogen level in the tank as \(E_0 = 1.6\)MWh and \(H_0 = 5\)Nm\(^3\) respectively. Other technical-economic parameters of the BESS and HESS are tabulated in Table \ref{sim_params}. A day is divided into 48 time slots, i.e., each time slot is equivalent to 30 minutes. 
\begin{table}[h!]
	\centering
	\caption{BESS and HESS Simulation Parameters.}
	\begin{adjustbox}{width = \columnwidth, center}
	\begin{tabular}{l l}%			  
		\hline
		ESS & Parameter \& Value\\			
		\hline
		\hline
\multirow{3}{*}{BESS}& \(E_{nom}=\)2MWh, \(P_{max}=102\)kW,  \(DoD=\)80\%\\ 
		&\(E_{min}=\)0.1MWh, \(E_{max}=\)1.9MWh, \(L_c=\)3650\\ 
		&\(C_b^{ca}\)=\pounds210000, \(\eta_{c,t}=\eta_{d,t}=\)98\%\ \\		 
		\hline 
\multirow{3}{*}{HESS}& \(H_{min}=\)2Nm\(^3\), \(H_{max}=\)10Nm\(^3\), \(P_{max}^{el}=\)3kW \\ 
		&\(P_{max}^{fc}=\)3kW, \(\eta_{fc,t}=\)50\%, \(\eta_{el,t}=\)90\%\\ 
		&\(L_{fc}=L_{el}= 30000\)h, \(r_{fc,t}=\)0.23Nm\(^3\)/kWh\\
		& \(r_{el,t}=\)1.32kWh/Nm\(^3\), \(C_{el}^{om}=C_{fc}^{om}\)=\pounds 0.174/h\\
		&\(C_{el}^{ca}\)=\pounds 60000, \(C_{fc}^{ca}\)=\pounds 22000\\			
		\hline 
		\hline
	\end{tabular} 
	\end{adjustbox}
	\label{sim_params}
\end{table}

The actor and critic networks for each MADDPG agent are designed using hyper-parameters tabulated in Table \ref{hyper_params}. We use Rectified Linear Unit (ReLU) as an activation function for the hidden layers and the output of the critic networks. A Tanh activation function is used in the output layer of each actor network. We set the capacity of the replay buffer to be \(\mathbf{K} = 1\times 10^6\) and the maximum training steps in an episode to be \(T = 48\). The  Algorithm \ref{algorithm} is developed and implemented in Python using PyTorch framework \cite{paszke2019pytorch}. 
\begin{table}[t!]
		\centering
		\caption{Hyper-parameters for each Actor and Critic Network.}
		\begin{adjustbox}{width = \columnwidth, center}
		\begin{tabular}{l l l}%			  
			\hline
			Hyper-parameter& Actor Network & Critic Network\\			
			\hline
			\hline
			Optimizer & Adam & Adam \\
			Batch size & 256  & 256\\			
			Discount factor & 0.95  & 0.95\\ 	
			Learning rate & $1\times 10^{-4}$ & $3\times10^{-4}$\\		
			No. of hidden layers & 2 & 2\\	
			No. of neurons & 500 & 500\\ 			 
			\hline 
			\hline
		\end{tabular} 
		\end{adjustbox}
		\label{hyper_params}
\end{table}

\subsection{Benchmarks}

We verify the performance of the proposed MADDPG algorithm by comparing it with other three bench-marking algorithms:
\begin{itemize}
    \item Rule-based (RB) algorithm: This is a model-based algorithm which follows the standard  practise of wanting to meet the energy demand of the SEN using the RES generation without guiding the  operation of BESS, HESS and flexible demands towards periods of low/high electricity price to save energy costs. In the event that there is surplus energy generation, the surplus is first stored in the short-term BESS, followed by the long-term HESS and any extra is sold to the main grid. If the energy demand exceeds RES generation, the deficit is first provided by the BESS followed by the HESS and then the main grid.  
    \item DQN algorithm: As discussed in Section \ref{sec:MADRL}, this is a value-based DRL algorithm, which intends to optimally schedule the operation of the BESS, HESS and flexible demand using a single agent and a discretised action space.
    \item DDPG algorithm: This is a policy-based DRL algorithm, which intends to optimally schedule the operation of the BESS, HESS and flexible demand using a single agent and a continuous action space as discussed in Section \ref{sec:MADRL}.
\end{itemize}

\subsection{Algorithm Convergence}
\begin{figure}[t!]
    \centering
    \includegraphics[width=0.48\textwidth]{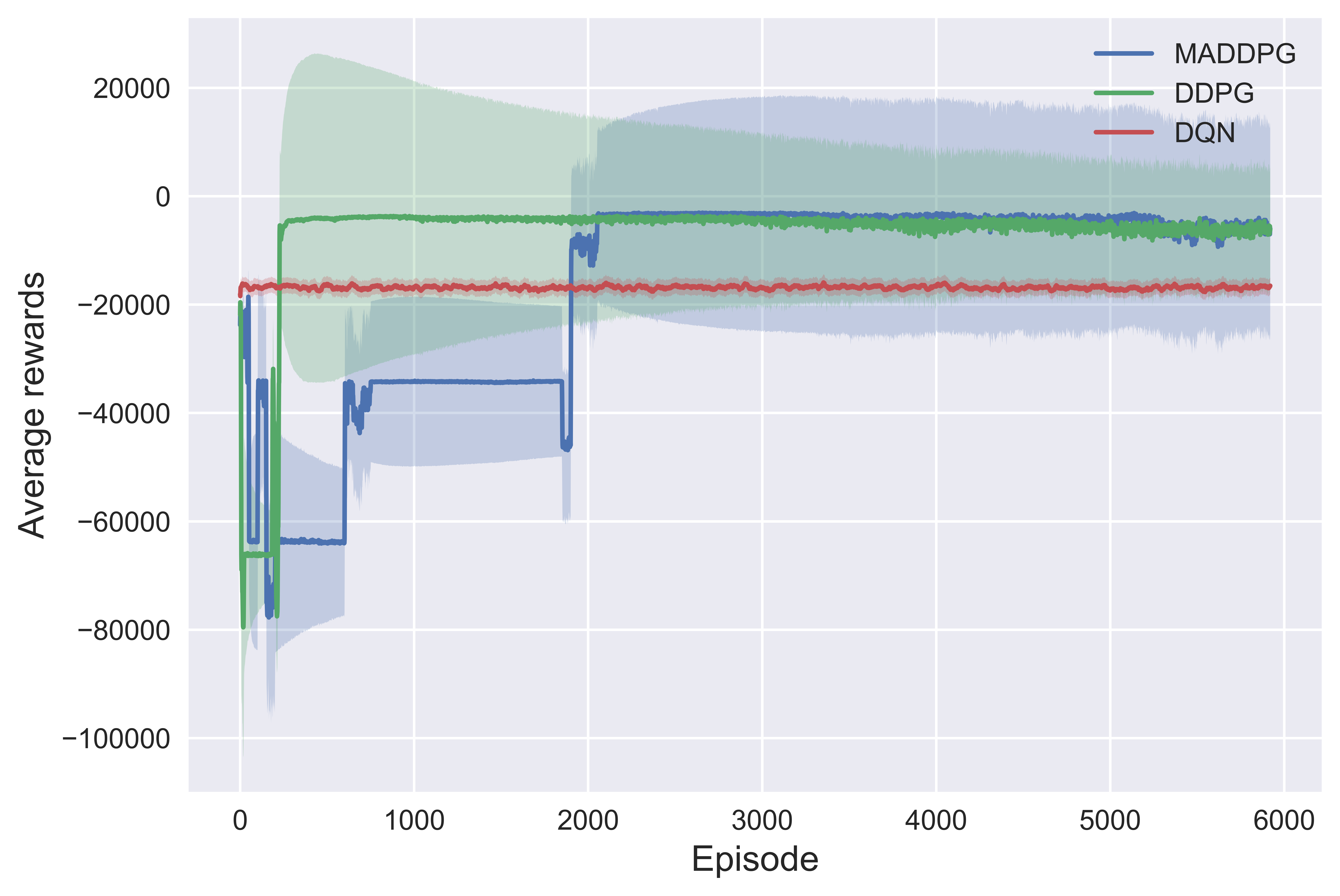}
    \caption{Training processes of the DQN, DDPG and MADDPG algorithms}
    \label{training_rewards}
    \vspace{-\baselineskip}
\end{figure}

We analyse the convergence of the  MADDPG algorithm by training the agents with 5900 episodes, with each episode having 48 training steps.  In Fig. \ref{training_rewards}, the average rewards obtained for each episode are plotted against the episodes and compared to the DRL-based bench-marking algorithms. 

As shown in Fig. \ref{training_rewards}, all algorithms achieve convergence after 2000 episodes. The DQN reach convergence faster than MADDPG and DDPG due to the DQN's discretised and low-dimensional action space, making the determination of the optimal scheduling policy relatively easier and quicker than the counterpart algorithms with continuous and high-dimensional action spaces. As a discretised action space cannot accurately capture the complexity and dynamics of the SEN energy management, the DQN algorithm converges to the worst optimal policy given by the lowest average reward value (-16572.5). On the other hand, the MADDPG algorithm converges to a high average reward value (-6858.1), which is slightly higher than the reward value (-8361.8) for the DDPG, mainly due to enhanced cooperation between the operation of the controlled assets.   

\begin{figure}[t!]
    \centering
    \includegraphics[width=0.49\textwidth]{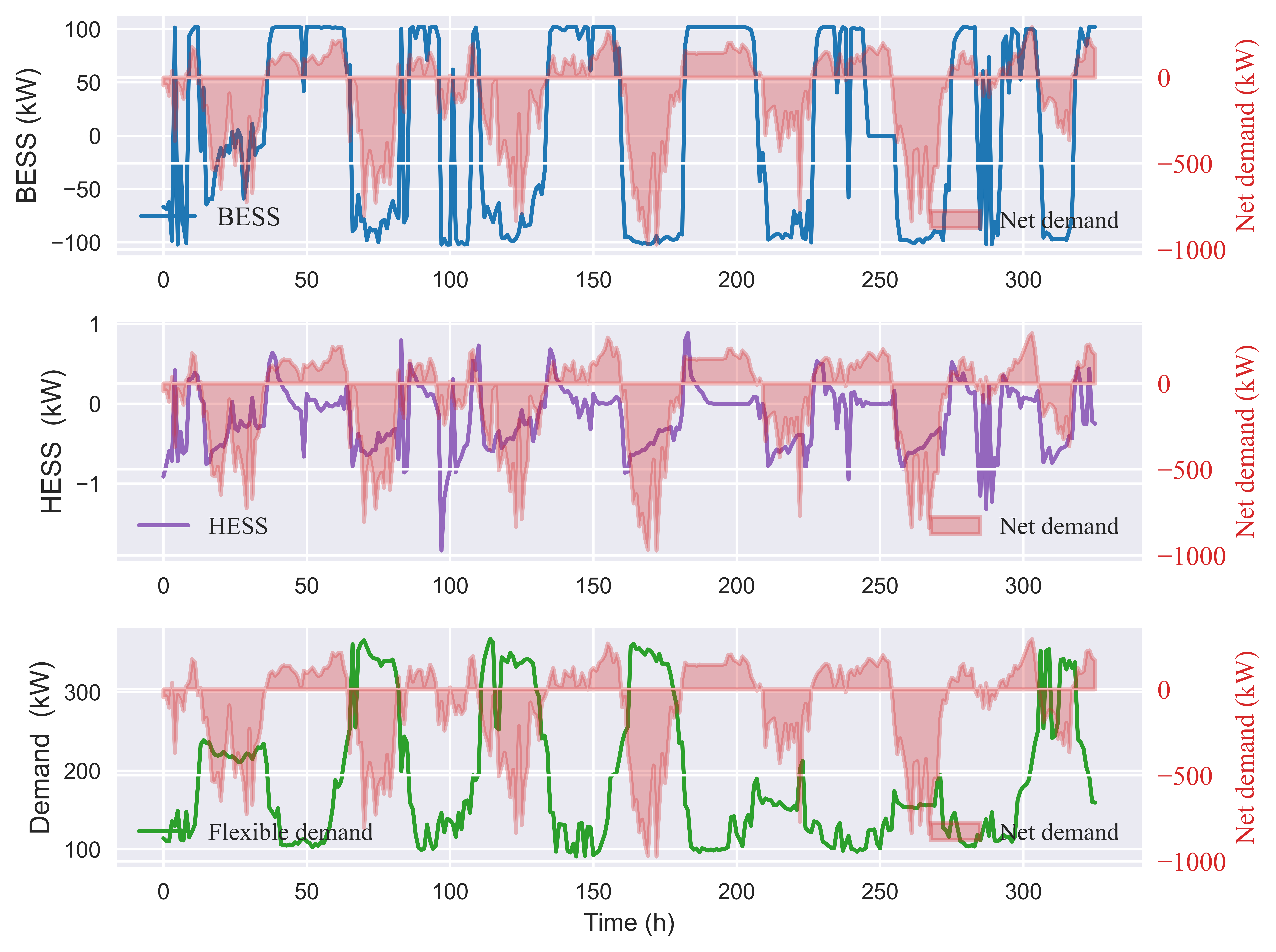}
    \caption{Control action results (for a 7 day-period) by BESS, HESS and Flexible demand agents with response to net demand.}
    \label{net_demand_actions}
    \vspace{-\baselineskip}
\end{figure}

\subsection{Algorithm Performance}
In this section, we demonstrate the effectiveness of the proposed algorithm for optimally scheduling the BESS, HESS and the flexible demand to minimise the energy and environmental costs. Fig. \ref{net_demand_actions} shows the scheduling results with response to the SEN net demand for a period of 7 days, i.e., \(T= 336\) hours. As shown in  Fig. \ref{net_demand_actions}, the BESS and HESS  accurately charge (negative power) and discharge (positive power) whenever the net demand is negative (i.e., RES generation exceeds energy demand) and positive (i.e., energy demand exceeds RES generation) respectively. Similarly, the scheduled demand is observed to be high and low whenever the net demand is negative and positive respectively. 

\begin{figure}[t!]
    \centering
    \includegraphics[width=0.49\textwidth]{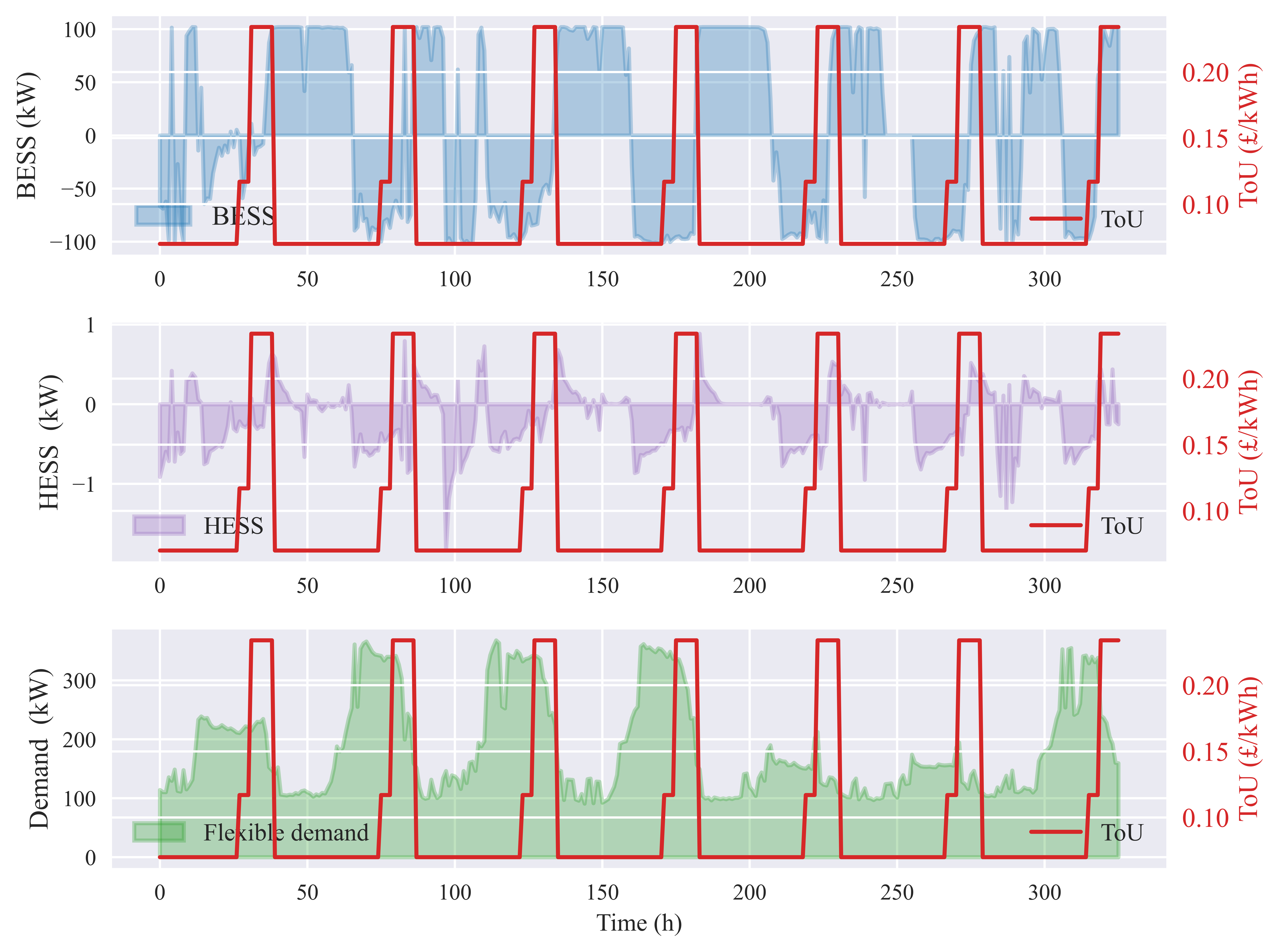}
    \caption{Control action results (for a 7 day-period) by BESS, HESS and Flexible demand agents with response to ToU.}
    \label{ToUactions}
    \vspace{-\baselineskip}
\end{figure}

Fig. \ref{ToUactions} shows that in order to minimise the multi-objective function given by \(\textbf{P1}\), the algorithm prioritises the flexible demand agent to aggressively respond to price changes compared to the BESS and HESS agents. As shown in Fig. \ref{ToUactions}, the scheduled demand reduces sharply whenever the electricity price is the highest and increases when the price is lowest compared to the actions by the BESS and HESS. 

Together, Fig. \ref{net_demand_actions} and Fig. \ref{ToUactions} demonstrate how the algorithm allocates different priorities to the agents to achieve a collective goal, which  is to minimise carbon costs, energy and operational costs. In this case, the BESS and HESS agents are trained to response more aggressively to changes in energy demand and generation, and maximise the benefits thereof like minimum carbon emissions. On the other hand, scheduling the flexible demand guides the SEN towards low energy costs.

\begin{table}[t!]
		\centering
		\caption{Cost Savings and Carbon Emissions for Different SEN Models.}
		\begin{adjustbox}{width = \columnwidth, center}
		\begin{tabular}{l| c c c c c}%			  
			\hline
			\textbf{Models} & \textbf{Proposed} & No BESS & No HESS & No Flex. Demand & No assets\\			
			\hline
			\hline
			BESS & \checkmark & $\times$ & \checkmark & \checkmark & $\times$\\			
			HESS & \checkmark  & \checkmark & $\times$ & \checkmark & $\times$\\ 			
			Flex. Demand & \checkmark & \checkmark & \checkmark & $\times$ & $\times$\\
			\hline
			\hline
			\textbf{Cost Saving} (£) &1099.60 & 890.36& 1054.58 & 554.01& 451.26\\
			\hline
			\textbf{Carbon Emission} (kgCo\(_2\)e)& 265.25 & 1244.70 & 521.92& 1817.37 & 2175.66\\
			\hline 
			\hline
		\end{tabular} 
		\end{adjustbox}
		\label{benefitsofBH}
\end{table}

\subsection{Improvement in Cost Saving and Carbon Emission}
To demonstrate the economic and environmental benefits of integrating the BESS and the HESS in the SEN, the MADDPG algorithm was tested on  different SEN models as shown in Table \ref{benefitsofBH}. The SEN models differ based on whether the SEN has any of the controllable assets; BESS, HESS and flexible demand or not. For example, the SEN model which only has HESS and flexible demand as controllable assets is denoted as `No BESS'. The total cost savings and carbon emissions for each model were obtained as a sum of the cost savings and carbon emissions obtained half-hourly for 7 days.

As shown in Table \ref{benefitsofBH}, integrating BESS and HESS in the SEN as well as scheduling the energy demand achieves the highest cost savings and reduction in carbon emission. For example, the cost savings and carbon emissions are 23.5\% and 78.69\% higher and lower respectively than those for the SEN model without BESS (i.e., the `No BESS' model), mainly due to improved RES utilisation for the proposed SEN model. 

\subsection{Improvement in RES Utilisation}
To demonstrate improvement in RES utilisation as a result of integrating the BESS and the HESS in the SEN as well as scheduling energy demand, we use self-consumption and self-sufficiency as performance metrics. Self-consumption is defined as a ratio of RES generation used by the SEN (i.e., to meet the energy demand and to charge the BESS and HESS) to the overall RES generation \cite{luthander2015photovoltaic}. Self-sufficiency is defined as the ratio of the energy demand that is supplied by the RES, BESS and HESS to the overall energy demand \cite{long2018peer}. 

 Table \ref{benefitsofSS} shows that integrating the BESS and the HESS in the SEN as well as scheduling energy demand improves RES utilisation. Overall, the proposed SEN model achieved the highest  RES utilisation with 59.6\% self-consumption and 100\% self-sufficiency. This demonstrates the potential of integrating HESS in future SENs for absorbing more RES, thereby accelerating the rate of power system decarbonisation. 
 
\begin{table}[t!]
		\centering
		\caption{Self-consumption and self-sufficiency for Different SEN Models.}
		\begin{adjustbox}{width = \columnwidth, center}
		\begin{tabular}{l| c c c c c}%			  
			\hline
			\textbf{Models} & \textbf{Proposed} & No BESS & No HESS & No Flex. Demand & No assets\\			
			\hline
			\hline
			BESS & \checkmark & $\times$ & \checkmark & \checkmark & $\times$\\			
			HESS & \checkmark  & \checkmark & $\times$ & \checkmark & $\times$\\ 			
			Flex. Demand & \checkmark & \checkmark & \checkmark & $\times$ & $\times$\\
			\hline
			\hline
			\textbf{Self-consumption} &59.6\% & 48.0\%& 39.2\% & 46.0\%& 50.0\%\\
			\hline
			\textbf{Self-sufficiency} &100\% & 85.3\%& 95.2\% & 78.8\%& 73.4\%\\
			\hline 
			\hline
		\end{tabular}
		\end{adjustbox}
		\label{benefitsofSS}
\end{table}

\subsection{Algorithm Evaluation}
\begin{figure}[t!]
	\centering
	\begin{subfigure}[b]{0.4\textwidth}
	\centering
	\includegraphics[width=\textwidth]{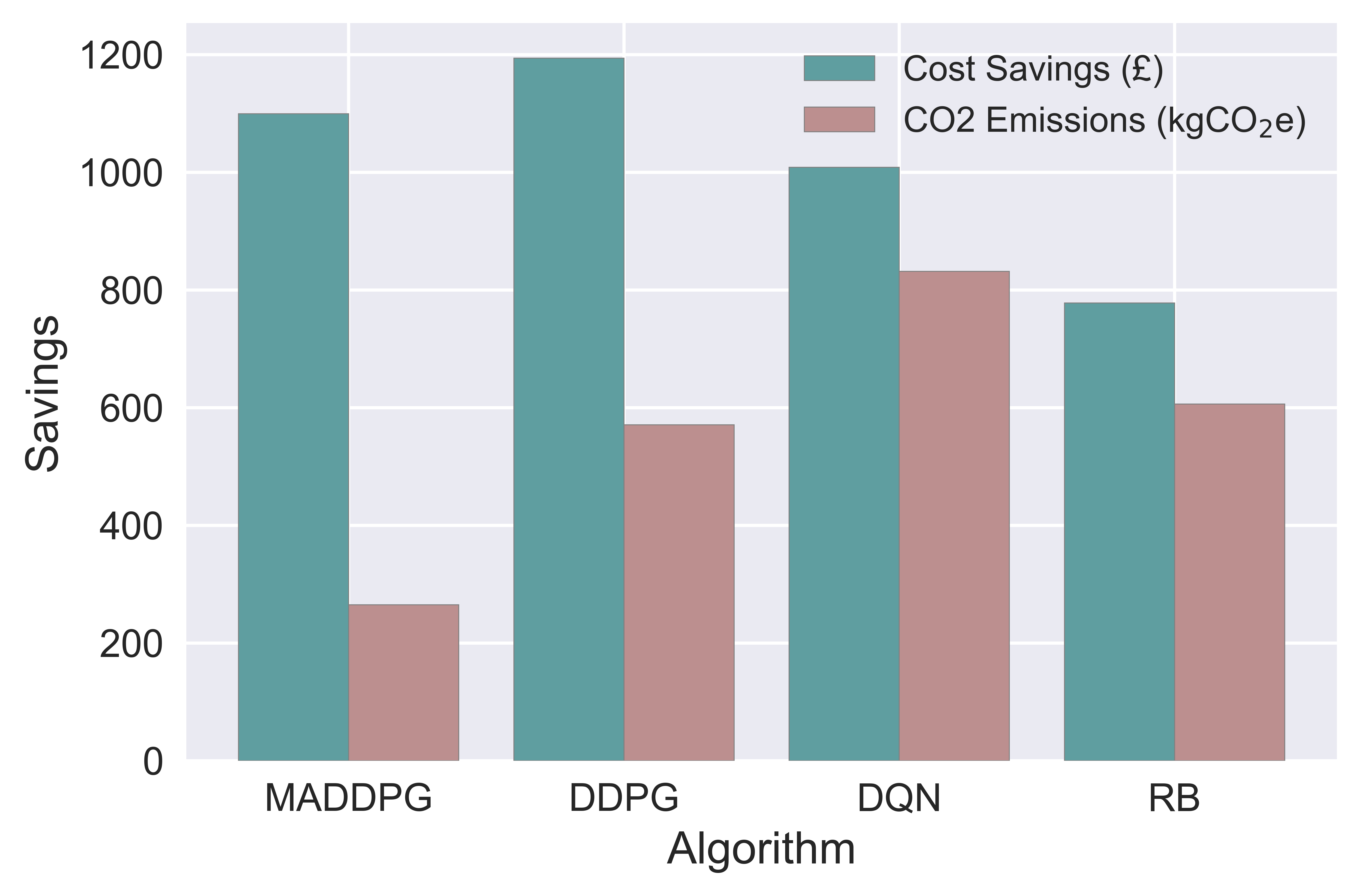}
	\caption{}
	\label{costs}
	\end{subfigure}
	\hfill
	\begin{subfigure}[b]{0.4\textwidth}
	\centering
	\includegraphics[width=\textwidth]{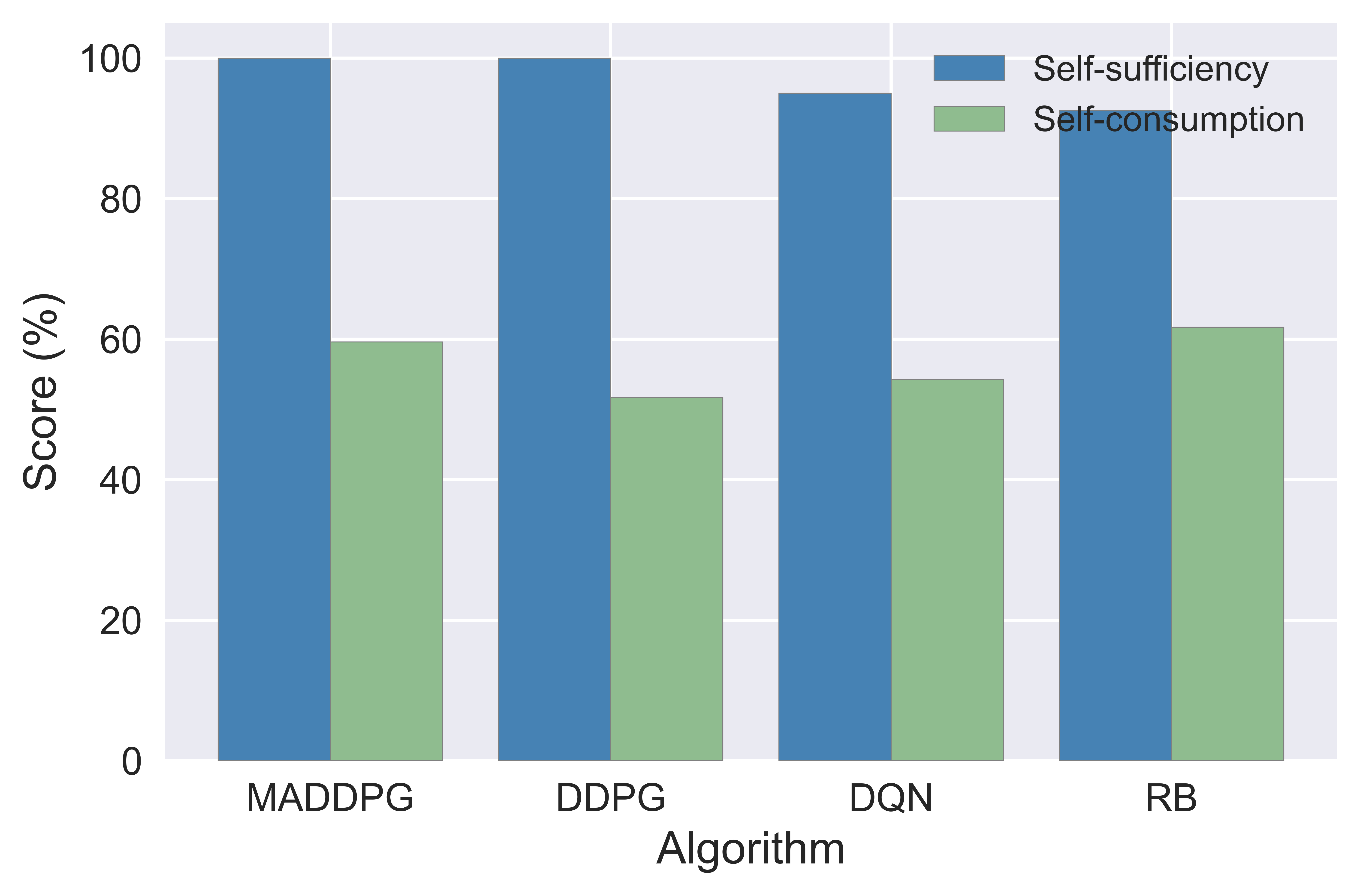}
	\caption{}
	\label{self_s}
	\end{subfigure}
	\caption{Performance of the MADDPG algorithm compared to the bench-marking algorithms for: (a) cost savings and carbon emissions, (b) self-consumption and self-sufficiency.}
	\label{bench_making}
\end{figure}

Performance of the proposed MADDPG algorithm was evaluated by comparing it to the  bench-marking algorithms for cost savings, carbon emissions, self-consumption and self-sufficiency  as shown in Fig. \ref{bench_making}.
The MADDPG algorithm obtained the most stable and competitive performance in all the performance metrics considered, i.e., cost savings, carbon emissions, self-consumption and self-sufficiency. This is mainly due to its multi-agent feature, thereby ensuring a better learning experience of the environment. For example, the MADDPG improved the cost savings and reduced the carbon emissions by 41.33\% and 56.3\% respectively relative to the RB approach. The rival DDPG algorithm achieved the highest cost savings at the expense of carbon emissions and self-sufficiency. As more controllable assets are expected in future SENs due to the digitisation of power systems, multi-agent based algorithms are therefore expected to play a key energy management role.

\subsection{Sensitivity Analysis of Parameter $\alpha_d$ }
\begin{figure}[t!]
    \centering
    \includegraphics[width=0.4\textwidth]{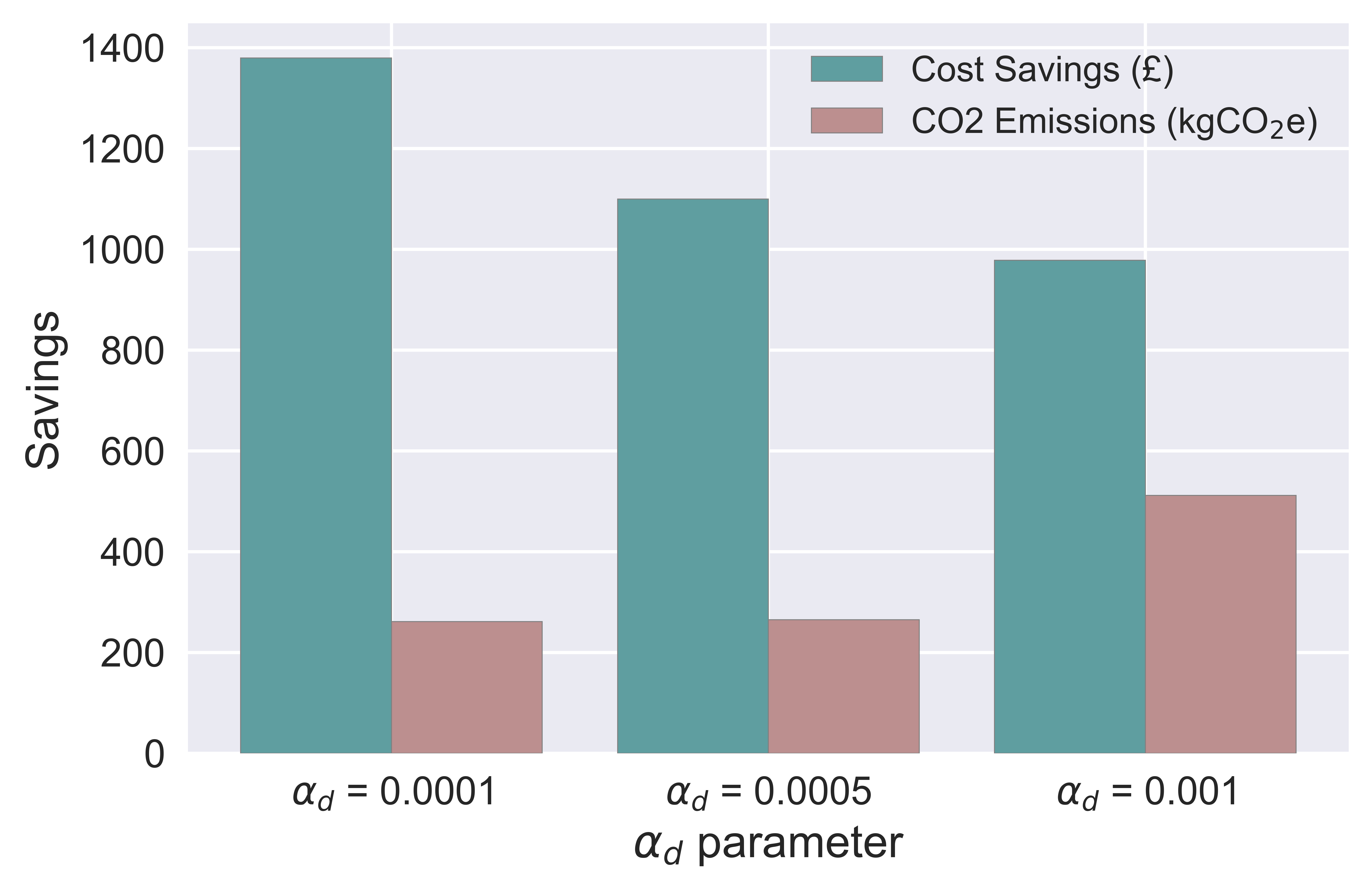}
    \caption{Cost savings and carbon emissions for different $\alpha_d$ parameters.}
    \label{parameter_alphas}
    \vspace{-\baselineskip}
\end{figure}

 The parameter \(\alpha_d\) quantifies  the amount of flexibility  to reduce the energy demand. A lower value of \(\alpha_d\) indicates that less attention is paid to the inconvenience cost and a larger share of the energy demand can be reduced to minimise the energy costs. A higher value of \(\alpha_d\) indicates that high attention is paid to the inconvenience cost and the energy demand can be hardly reduced to minimise the energy costs. With the change in \(\alpha_d\) values, the cost saving and carbon emission results are compared in Fig. \ref{parameter_alphas}.
 
 As shown in Fig. \ref{parameter_alphas}, the cost savings and carbon emissions reduce and increase respectively as  \(\alpha_d\) takes values from 0.0001 to 0.001, which means that the energy demand's sensitivity to price reduces with increased inconvenience levels as given by (\ref{inc_cost}). Thus, having an energy demand which is sensitive to electricity price is crucial for reducing carbon emissions and  promoting the use of RES. 
\section{Conclusions}\label{conclusion}
In this paper, we investigated the problem of minimising energy costs and carbon emissions as well as increasing renewable energy utilisation in a smart energy network (SEN) with BESS, HESS and schedulable energy demand. A multi-agent deep deterministic policy gradient algorithm  was proposed as a real-time control strategy to optimally schedule the operation of the BESS, HESS and schedulable energy demand while ensuring that the operating constraints and time-coupled storage dynamics of the BESS and HESS are achieved. Simulation results based on real-world data showed increased cost savings, reduced carbon emissions and improved renewable energy utilisation with the proposed algorithm and SEN. On average, the cost savings and carbon emissions were 23.5\% and 78.69\% higher and lower respectively with the proposed SEN model than baseline SEN models. The simulation results also verified the efficacy of the proposed algorithm to manage the SEN outperforming other bench-marking algorithms including DDPG and DQN algorithms. Overall, the results have shown great potential for integrating HESS in SENs and using self-learning algorithms to manage the operation of the SEN. 

\section*{Acknowledgement}
This work was supported by the Smart Energy Network Demonstrator project (grant ref. 32R16P00706) funded by ERDF and BEIS. This work is also supported by the EPSRC EnergyREV project (EP/S031863/1).

 \bibliographystyle{elsarticle-num} 
 \bibliography{cas-refs}





% \end{thebibliography}
\end{document}